# The Interface Conditions for Pressures at Oil-water Flood Front in the Porous Media Considering Capillary Pressure

Xiaolong Peng, Fei Mo, Zhimin Du

(State Key Laboratory of Oil and Gas Reservoir Geology and Development Engineering, Southwest Petroleum University, Chengdu. 610500)

## ABSTRACT

Flood front is the jump interface where fluids distribute discontinuously, whose interface condition is the theoretical basis of a mathematical model of the multiphase flow in porous medium. The conventional interface condition at the jump interface is expressed as the continuous Darcy velocity and fluid pressure (named CPVCM). This paper has inspected it via the studying the water-oil displacement in one dimensional reservoir with considering capillary pressure but ignoring the compressibility and gravity. It is proved theoretically that the total Darcy velocity and total pressure (defined by Antoncev etc.), instead of the Darcy velocities and pressures of water and oil, are continuous at the flood front without considering the compressibility of fluid and porous media. After that, new interface conditions for the pressures and Darcy velocity of each fluid are established, which are collectively named as Jump Pressures and Velocities Conditions Model (JPVCM) because the model has shown the jump pressures and jump Darcy velocities at the flood front. Finally, three application-examples are proposed and the results show JPVCM is more reasonable than CPVCM.

**Keywords:** Flood front; Jump Condition; Two phase; numerical flux; mathematical model

| Nomenclature | | | |
|---|---|---|---|
| $f$ | fractional flow, dimensionless | $t$ | time, T |
| $K$ | absolute permeabillity, $L^2$ | $x$ | distance, L |
| $k_c$ | relative permeability, dimensionless | $\Gamma$ | Interface, symbol. |
| $\phi$ | porosity, dimensionless | $\lambda$ | mobility, $LM^{-1}T$ |
| $\mu$ | viscocity, $L^{-1}MT^{-1}$ | **Subscripts** | |
| $s$ | saturation, dimensionless; | $w$ | water |
| $S$ | saturation, dimensionless; | $n$ | non-wet fluid, or oil |
| $P$ | total pressure, $L^{-1}MT^{-2}$ | $o$ | oil |
| $P_\ell$ | pressure of fluid $\ell$, $L^{-1}MT^{-2}$ | $t$ | total |
| $P_c$ | capillary pressure, $L^{-1}MT^{-2}$ | - | limit on the upstream side |
| $v$ | Darcy velocity, $ML^{-1}$ | + | limit on the downstream side |
| | | in | inflow |
| | | out | outflow |

## 1. Introduction

The seepage differential equation is constructed on the basis of seepage mechanics and continuum theory, including medium continuity and fluid continuity(Bear, 2013). However, discontinuity in the reservoir is pervasively found, such as interlayer differences(Sattler, ImmenhauserA, HILLGÄRTNER, & Steban, 2005), oil-gas-water contact, and water or gas injection front(Glimm, Grove,Li, et.tal.,1998).. Boundary- or bottom-aquifer flood front (Bakker, 2002), interface of matrix and fractures(Cooper, Kelly, 1969; Moench, 1984), joint surface of wellbore and reservoir(Brohi, 2011), interface of the hydraulic fractures and micro-porosities(Pan, Oldenburg, 2014).And the study of the multiphase flow across that discontinuous is known as the Neumann problem in mathematics (Raeini, Blunt, Bijeljic, 2012; Cheng, Liggett, Lee, 198;). The whole reservoir can be divided into several continuous sub-regions by all the discontinuous interfaces, and within each of them, the differential equations of multiphase flow in porous media can be established. But in order to build a complete reservoir percolation mathematical model, the interface condition, namely coupling conditions of the flood front, require to be determined.

In accordance with the conventional documents about the multiphase seepage mechanics, the first condition states that normal component of the velocity must be continuous across the discontinuous interface. Water pressures on both sides are approximately equal in value (Denbigh, 1981; Szymkiewicz, 2012) as shown in Formula (1),

$$\begin{cases} v_\ell^- \cdot N_\Gamma = v_\ell^+ \cdot N_\Gamma & (\ell = o, w) \\ P_\ell^- = P_\ell^+ & (\ell = o, w) \end{cases} \tag{1}$$

The above conditions are widely applied in related researches on multiphase flow in porous medium (Szymkiewicz, 2012; Valdes-Parada, Espinosa-Paredes, 2005; Brenner, Cances, Hilhorst, 2013; Szymkiewicz, 2012; Hassanizadeh, Gray, 1989; Doster, Hilfer, 2011; Ohlberger, Schweizer, 2007; Mozolevski, Schuh, 2013; Huber, Helmig, 2000). However, we hold that at the flood front, the velocity and pressure of each phase fluid is not continuous. We have studied the interface condition of the flood front in the case of ignoring capillary pressure, which is called Jump Velocity Condition Model (JVCM) for the sake of easily distinguishing in a paper(Peng,et.al.2016，2007):

$$\begin{cases} v_n^- + v_w^- = v_n^+ + v_w^+ = v_t \\ v_\ell^- - v_\ell^+ = (f_\ell^- - f_\ell^+) v_t & (\ell = w, n) \\ P^- = P^+ \end{cases} \tag{2}$$

Where:

$$f_\ell \triangleq \frac{\lambda_\ell}{\lambda_n + \lambda_w} \tag{3}$$

In the formula above, $f_w^1$ and $f_w^2$ respectively represent the fractional function of water within the interval of $[x_1, x_\Gamma]$ and $(x_1, x_\Gamma]$ separately, and

$$\lambda_\ell \triangleq \frac{k_{r\ell}}{\mu_\ell} \tag{4}$$

Equation (2) implies that there is usually a velocity jump at the flood front because generally $f_\ell^- \neq f_\ell^+$. Nevertheless, based on the previous research results, a further consideration over the effect of capillary pressure is taken in this article to establish interface condition model of flood front. At first, proof by contradiction is adopted to prove that not only the seepage velocity of each phase at flood front is characterized by seepage discontinuity, so is the pressure. Then, as per the oil-water two phase seepage differential equation, interface condition on abrupt interface is concluded from the weak solution form in the scope of reservoir region (including discontinuous interface) (Chen, 1992; Arbogast, 1992; DiBenedetto, 1981; Baber, Mosthaf, Flemisch, et.al., 2012). Limited to the length of paper, gravity effect is excluded. Let define the expression below.

$$F^- \triangleq \lim_{\substack{\varepsilon>0 \\ \varepsilon\to 0}} F(x-\varepsilon), \quad F^+ \triangleq \lim_{\substack{\varepsilon>0 \\ \varepsilon\to 0}} F(x+\varepsilon) \tag{5}$$

Where, F presents any function of x.

## 2. Demonstrating the Inequality of Pressures on Both Side of the Flood Front

Considering the one-dimensional oil-water two phase seepage, and ignoring the compressibility of rocks and fluids, and overpassing gravity, flood front is symbolized by $\Gamma$, which separates the reservoir into two continuum regions $\Omega_1$ and $\Omega_2$, and the whole reservoir region is expressed as $\Omega_1 \cup \Omega_2 \cup \Gamma$. The flow in each of the sub continuous region meets Mustkat equations (Muskat, 1937):

$$\begin{cases} \nabla \cdot v_n + \phi \dfrac{\partial s_n}{\partial t} = 0 \\ \nabla \cdot v_w + \phi \dfrac{\partial s_w}{\partial t} = 0 \end{cases} \quad x \in \{\Omega_1 \cup \Omega_2 \setminus \Gamma\} \tag{6}$$

Motion equation

$$v_\ell = -K\lambda_\ell \nabla P_\ell \tag{7}$$

Where $\lambda_\ell \triangleq \dfrac{k_{r\ell}}{\mu_\ell}$.

Auxiliary equation

$$s_w + s_n = 1 \tag{8}$$

$$P_c = P_n - P_c \tag{9}$$

**Theorem.1**: *As for water driving oil in the porous medium described in equations (6)-(9), it has been known that the oil-water flood front is located at $x_\Gamma$; saturation on each side of the flood front $\Gamma$ is $s_\ell^-, s_\ell^+ (\ell = n, w)$, and $s_w^- = s_{wf}^- \neq s_w^+ = s_{wi}^+$; permeability is expressed as $K$; relative permeability curve*

*is marked as* $k_{r\ell}(s_\ell)$; *capillary pressure* $P_c = P_c(s_w)$, *which is a strictly monotone function of water saturation. Therefore, it comes the conclusion that oil pressure and water pressure are not continuous at the flood front interface.*

We adopt the method of proof by contradiction to demonstrate **Theorem.1**. So we have the following hypothesis.

**[H2.1]** *Assume the conclusions of **Theorem.1** are false, so that, oil pressures ,as well water pressures, on both sides of the flood front are equal.*

Because it is known that $P_\ell(\ell = n, w)$ is continuous within $\Omega_1$ and $\Omega_2$, and so it is assumed at $\Gamma$, then $P_\ell(\ell = n, w)$ is continuous field in $\Omega_1 \bigcup \Omega_2 \bigcup \Gamma$.

According to the definition of capillary pressure, $P_c = P_n - P_w$, the equation that $P_c^- = P_c^+$ can be deduced. In light of the known conditions, $P_c$ is a strictly monotone function of saturation, so that

$$S_w^- = P_c^{-1}(S_w^-), S_w^+ = P_c^{-1}(S_w^+) \text{ and } S_w^- = S_w^+ \tag{10}$$

Where $P_c^{-1}$ represents the inverse of capillary pressure function.

Hence, from the known condition that $S_w^- \neq S_w^+$, we can find that the deduced conclusion contradicts with the known condition. Therefore, the hypothesis **[H2.1]** is untenable and the **Theorem.1** is adequately demonstrated.

## 3. The Interface Condition for Fluid Pressures at the Flood Front.

We consider the oil-water two phase flow in one-dimensional reservoir, and capillary pressure, and ignore gravity, Water driving oil front is located at $x_\Gamma$, take two adjacent infinitesimal cells on both sides of front $\Gamma$, respectively expressed as $\Omega_1$ and $\Omega_2$, and we get the equation, $\Gamma = \overline{\Omega}_1 \bigcap \overline{\Omega}_2$. From the known center coordinates of cell $\Omega_1$ and cell $\Omega_2$, respectively presented as $x_1$ and $x_2$, there exists the relation $x_1 < x_\Gamma < x_2$, so does the pressure at $x_1$ and $x_2$ expressed as $P_\ell^1$ and $P_\ell^2$ $(\ell = o, w)$. Fluid saturations of the reservoir are known as Fig.1. Please note that the saturation within the cells is not constant, which means this case is more in line with the real situation of general reservoirs.

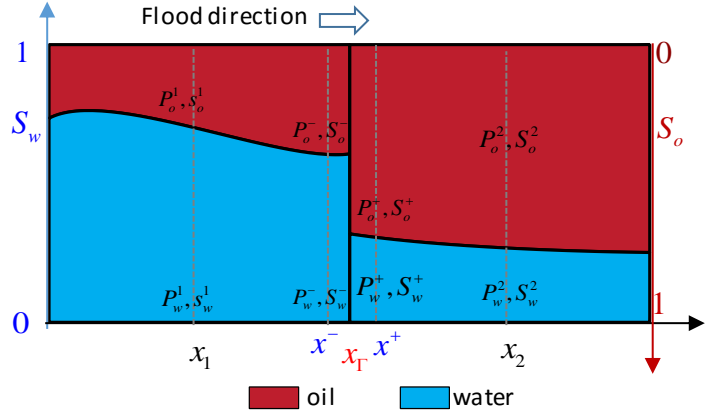

Fig. 1 Pressure and Saturation Distribution of Water-driving-oil Front

We might as well mark the saturation at $x_1$ as $s_\ell^1$, and $x_2$ as $s_\ell^2$; mark the pressure as $P_\ell^-$, saturation as $s_\ell^-$, when $x^- \triangleq \lim\limits_{\substack{\varepsilon>0 \\ \varepsilon \to 0}}(x_\Gamma - \varepsilon)$; and pressure as $P_\ell^+$, saturation as $s_w^+$, when $x^+ \triangleq \lim\limits_{\substack{\varepsilon>0 \\ \varepsilon \to 0}}(x_\Gamma + \varepsilon)$.

**3.1 Model the Coupling Boundary Condition for Pressures at Flood Front**

As for MUSKAT equations with discontinuous interface, in light of the seepage mechanics theory, it can be converted to the ones in form of weak solution in Sobrov space (Arbogast, 1992; Chen, 2001; Chen, Huan, Ma, 2006; Costa, Oliveira, Baliga, etal. 2004):

$$\begin{cases} \int_{x1}^{x2}\left(\dfrac{\partial v_n}{\partial x} + \phi \dfrac{\partial s_n}{\partial t}\right) = 0 \\ \int_{x1}^{x2}\left(\dfrac{dv_w}{dx} + \phi \dfrac{\partial s_w}{\partial t}\right) = 0 \end{cases} \quad x \in \{\Omega_1 \cup \Omega_2 \cup \Gamma\} \qquad (11)$$

Add together the oil-water two phase continuous formulates in Equation Set (11), make $v_t \triangleq v_n + v_w$, $\lambda_t \triangleq \lambda_n + \lambda_w$, and $v_t$ is deduced to be a constant, namely:

$$v_t = -K\lambda_t\left(\dfrac{dP_n}{dx} - f_w \dfrac{dP_c}{dx}\right) \quad x \in \{\Omega_1 \cup \Gamma \cup \Omega_2\} \qquad (12)$$

Divide by $-\lambda_t K$ at both side of the equation (12), do finite integral in the scope of $[x_1, x_2]$, and suppose $\varepsilon > 0$, as well as $\varepsilon \to 0$:

$$v_t = \dfrac{\int_{x_1}^{x_2}\dfrac{dP_n}{dx}dx - \int_{x_1}^{x_2}\left(f_w \dfrac{\partial P_c}{\partial x}\right)dx}{-\int_{x_1}^{x_\Gamma}\dfrac{1}{\lambda_t K}dx - \int_{x_\Gamma}^{x_2}\dfrac{1}{\lambda_t K}dx}$$

$$= \frac{\left(P_{n2} - \int_0^{s(x_2)} f_w^1(\xi)\frac{dP_c^1(\xi)}{d\xi}d\xi\right) - \left(P_{n1} - \int_0^{s(x_1)} f_w^2(\xi)\frac{dP_c^2(\xi)}{d\xi}d\xi\right)}{-\int_{x_1}^{x_\Gamma} \frac{1}{\lambda_t^1 K^1}dx - \int_{x_\Gamma}^{x_2} \frac{1}{\lambda_t^1 K^1}dx} \quad (13)$$

where $P_c^1, P_c^2$ respectively symbolize the capillary pressure curve within the interval of $[x_1, x_\Gamma]$, $(x_\Gamma, x_2]$. With the reference of the methods raised by Antoncev, Chavent, etc.(Antoncev, 1972; Chavent, 1976; Arbogast, 1992; Chen, 2001), define the global pressure:

$$P = P_n - \int_0^s f_w(\xi)\frac{dP_c(\xi)}{d\xi}d\xi \quad (14)$$

Input the Equation (7) into (6), and obtain the following:

$$v_t = \frac{P_2 - P_1}{-\int_{x_1}^{x_\Gamma} \frac{1}{\lambda_t K}dx - \int_{x_\Gamma}^{x_2} \frac{1}{\lambda_t K}dx} \quad (15)$$

Next, we will prove the total pressure at the flood front to be continuous, i.e. $P^- = P^+$.

**Theorem.2:** *As for the problem of one-dimensional seepage, oil-water two phase percolation meets the MUSKAT equations shown in Equation (1) and the total pressure defined in Equation* (14) *if ignoring fluid and rock compressibility, then $P^+ = P^-$ in the discontinuous interface, with P continuing in the reservoir.*

Analysis: when $v_t = 0$, clearly $P^+ = P^-$. But to apply it in a more general situation where $v_t \neq 0$, Formula (15) is the main theory basis used for the proof.

Proof: according to Formula (15), there has the following relationship:

$$v_t = \frac{P^- - P_1}{-\int_{x_1}^{x_\Gamma} \frac{1}{\lambda_t K}dx} \quad (16)$$

Via Formula (16), get the following:

$$P_1 = P^- + v_t \int_{x_1}^{x_\Gamma} \frac{1}{\lambda_t K}dx \quad (17)$$

From Formula (15), directly get the following:

$$P_2 = P_1 - v_t \left(\int_{x_1}^{x_\Gamma} \frac{1}{\lambda_t K}dx + \int_{x_\Gamma}^{x_2} \frac{1}{\lambda_t K}dx\right) \quad (18)$$

Substitute Formula (17) into (18), and get:

$$P_2 = P^- - v_t \int_{x_\Gamma}^{x_2} \frac{1}{\lambda_t K}dx \quad (19)$$

From Formula (15), we get:

$$v_t = -\frac{P_2 - P^+}{\int_{x_\Gamma}^{x_2} \frac{1}{\lambda_t K} dx}$$

(20)

Namely:

$$P^+ = P_2 + v_t \int_{x_\Gamma}^{x_2} \frac{1}{\lambda_t K} dx$$

(21)

Substitute Formula (19) into (21) and get:

$$P^+ = P^-$$

(22)

Formula (22) shows that the left and right limit of the total pressure field $P$ at $x_\Gamma$ equates, and because $P_\ell$ and $P_c$ are continuous inside the sub regions of cell $\Omega_1$ and cell $\Omega_2$, so $P$ is a continuous field. Therefore, Theorem 2 is proved.

Deducing from the Theorem 2 and the definition of total pressure, the interface condition for oil pressure can be gotten:

$$P_n^- - P_n^+ = \int_0^{s_w^-} f_w^1(\xi) \frac{dP_c^1(\xi)}{d\xi} d\xi - \int_0^{s_w^+} f_w^2(\xi) \frac{dP_c^2(\xi)}{d\xi} d\xi$$

(23)

Substitute $P_c = P_n - P_w$ into Formula (23), and get the interface condition for water pressure at the flood front:

$$P_w^- - P_w^+ = -P_c^1(s_w^-) + P_c^2(s_w^+) - \int_0^{s_w^-} f_w^1(\xi) \frac{dP_c^1(\xi)}{d\xi} d\xi + \int_0^{s_w^+} f_w^2(\xi) \frac{dP_c^2(\xi)}{d\xi} d\xi$$

(24)

### 3.2 Model the Coupling Conditions for Fluid Velocities at Flood Front

By Equation (7), we can get:

$$v_w = f_w v_t + \frac{\lambda_n \lambda_w K}{\lambda_t} \frac{\partial P_c}{\partial x}$$

(25)

And:

$$v_n = f_n v_t - \frac{\lambda_n \lambda_w K}{\lambda_t} \frac{\partial P_c}{\partial x}$$

(26)

Therefore, Employing (15) and (25), we can get the interface condition for water velocity at flood front $\Gamma$

$$v_w^+ - v_w^- = \frac{(f_w^+ - f_w^-)(P_2 - P_1)}{-\int_{x_1}^{x_\Gamma} \frac{1}{\lambda_t K} dx - \int_{x_\Gamma}^{x_2} \frac{1}{\lambda_t K} dx} + \left( \frac{\lambda_n^+ \lambda_w^+ K^+}{\lambda_t^+} \frac{\partial P_c^+}{\partial x} - \frac{\lambda_n^- \lambda_w^- K^-}{\lambda_t^-} \frac{\partial P_c^-}{\partial x} \right)$$

(27)

Similarly, via equation (15) and (26), Interface condition for oil velocity at flood front $\Gamma$ can be obtained,

$$v_n^+ - v_n^- = \frac{(f_n^+ - f_n^-)(P_2 - P_1)}{-\int_{x_1}^{x_\Gamma} \frac{1}{\lambda_t K} dx - \int_{x_\Gamma}^{x_2} \frac{1}{\lambda_t K} dx} - \left( \frac{\lambda_n^+ \lambda_w^+ K^+}{\lambda_t^+} \frac{\partial P_c^+}{\partial x} - \frac{\lambda_n^- \lambda_w^- K^-}{\lambda_t^-} \frac{\partial P_c^-}{\partial x} \right) \quad (28)$$

When $f_w^- \neq 0, f_w^+ \neq 0$, Formula (27) can be rewritten as follows:

$$\left( v_w^- + \frac{\lambda_n^- \lambda_w^- K^-}{\lambda_t} \frac{\partial P_c^-}{\partial x} \right) : \left( v_w^+ + \frac{\lambda_n^+ \lambda_w^+ K^+}{\lambda_t} \frac{\partial P_c^+}{\partial x} \right) = f_w^- : f_w^+ \quad (29)$$

When $f_n^- \neq 0, f_n^+ \neq 0$, Formula (28) can be rewritten as follows:

$$\left( v_o^- - \frac{\lambda_o^- \lambda_w^- K^-}{\lambda_t} \frac{\partial P_c^-}{\partial x} \right) : \left( v_o^+ - \frac{\lambda_o^+ \lambda_w^+ K^+}{\lambda_t} \frac{\partial P_c^+}{\partial x} \right) = f_n^- : f_n^+ \quad (30)$$

## 4. Example

In this paper, 3 popular cases will be employed to analyze the pressure and velocity distribution at flood front and compare the results calculated by the new condition model and the conventional one.

### 4.1 Case 1. Piston-like Water-oil Displacement inside a Single Capillary

As shown in Fig 2, within a horizontal capillary of constant radius, water displaces oil in piston type, water and oil contact interface is expressed as $\Gamma$ with the coordinate $x_\Gamma$. Its capillary pressure function of water saturation is known. Water saturation on both sides are $s_w^1 (s_w^1 < 1, s_w^1 \to 1)$ and $s_w^2$ ($s_w^2 > 0, s_w^2 \to 0$), respectively. Flanked by $\Gamma$ are two points, A and B, whose pressures and saturations of the fluids are known. The water injection velocity is $v_t$. The task is to discover the pressure and velocity distribution of oil phase and water phase inside the capillary.

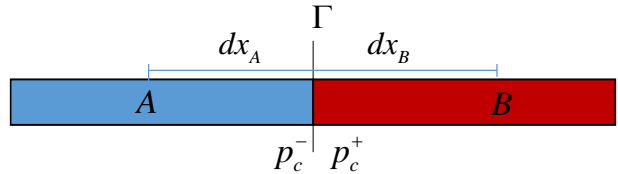

Fig.2 The Diagram of Piston-like Water-Oil Displacement within the Single Capillary

Table 1 Oil-water Relative Permeability Curve of the Capillary

| Water saturation | Water Relative Permeability | Oil Relative permeability |
| --- | --- | --- |
| 0 | 0 | 1 |
| 1 | 1 | 0 |

As there is only one capillary with constant radius, the capillary pressure curve is a horizontal line. That is to say, within the scope of the water saturation (0, 1), the value of capillary pressure is constant, which is marked as $P_c$ ($P_c > 0$ and $P_c(1) = 0$). Considering $s_w(x < x_\Gamma) \to 1$ on the side A of $\Gamma$ and $s_w(x > x_\Gamma) \to 0$ on the side B of $\Gamma$, therefore

$$\frac{dP_c}{ds_w} = 0 \quad (x \in [0, x_\Gamma) \cup (x_\Gamma, 1]) \tag{31}$$

Relative permeability curve can be expressed by two diagonal lines (refer to Table.1). so

$$f_w(s_w) = s_w \tag{32}$$

Moreover, because $\lim_{s_w^1 \to 1} P_c(s_w^1) = 0$, $\lim_{s_w^2 \to 0} P_c(s_w^2) = P_c$, so

$$p_o^A = P_w^A, P_o^B = P_w^B + P_c \tag{33}$$

Employing them and formula (23) and (24), obtains:

$$P_o^- - P_o^+ = P_c, P_w^- - P_w^+ = 0 \tag{34}$$

Substitute (31)-(32) into (27) and (28) and get:

$$v_w^+ - v_w^- = -v_w^- = -\frac{(P_w^B + P_c^B - P_w^A)}{\frac{dx_A}{\lambda_w K} + \frac{dx_B}{\lambda_o K}} = -v_t, v_o^+ - v_o^- = v_o^+ = -\frac{(P_w^B + P_c^B - P_w^A)}{\frac{dx_A}{\lambda_w K} + \frac{dx_B}{\lambda_o K}} = v_t \tag{35}$$

It is worth noticing that formula (34) and (35) show $P_w^- - P_o^+ = P_c$, $v_w^- = v_o^+$, which is consistent with the hydrodynamic experiment(Washburn, 1921; Rideal, 1922; Jánský, Tholin, Bonaventura, 2010; Masoodi, Languri, Ostadhossein, Fatt, 1956) but opposite to the conventional interface condition model that is described by equation(1).

Similar to the results from JPVCM as shown in Fig.3, the conventional interface conditions model also report that water pressure is continuous at the flood front $\Gamma$, so the water pressure field is the same as that in Fig.3. As for the oil pressure, according to equation (1), the point $(P_o^+, x^+)$ should be placed at Point.①, However according to the formula (9) and (37), it should be place at Point ②. Certainly, it is impossible to put one point at two different position. In order to avert this problem, a popular opinion is usually exposed that the fluids should modify its distribution automatically to submit the equation $P_c^- = P_c^+$. However, the example of water displacing oil in a single capillary is simple and clear enough to negate that opinion, because such water and oil distribution illustrated in Fig.2 really exist in the physical world and the fluid pressures have already been recognized or can be measured.

After obtained equation (34) and (35), it is turned to calculate the velocity field and pressure field in the entire capillary, which are illustrated in Fig.3 and Fig.4.

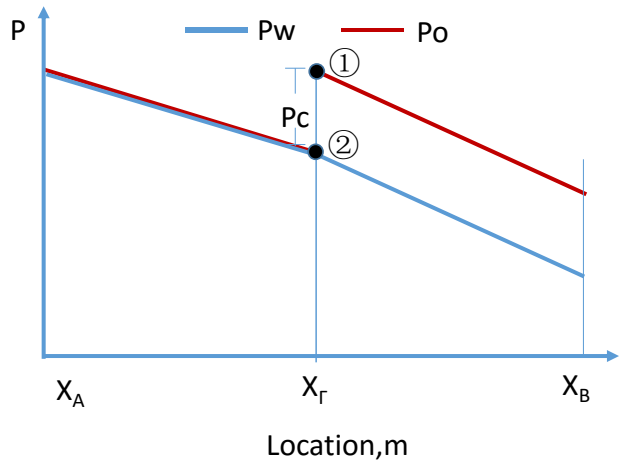
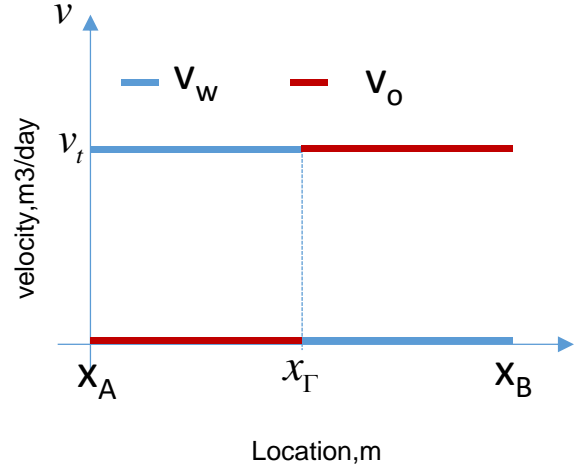

Fig.3 Pressure Distribution of Fluid in Capillary      Fig.4 Darcy Velocity of the Fluid in Capillary

**4.2 Case 2. The Piston-like Water-oil Displacement in Capillary Bundle of Different Radius**

As shown in Fig.5, the horizontal capillary bundle is of different sizes in radius(Dahle, Celia, and Hassanizadeh, 2005). Suppose that special experimental means are used to ensure the piston-like oil/water distribution. The fluids jump interface is symbolized as $\Gamma$, with its coordinates $x_\Gamma$. The water saturation on both sides of $\Gamma$, is 100% and 0 respectively，capillary pressure and relative permeability curve inside of the capillary are known (as shown in Table.2; Brooks, Corey, 1964.); A and B are two points respectively located in each side of $\Gamma$, where $P_w^A = 2$ MPa and $P_o^B = 1.9$ MPa. And the absolute permeability of capillary $K = 1$ D; the viscosity of water is 0.5cp and that of oil is 5cp; $dx_A = dx_B = 0.25$ m. The tasks are to calculate the interface condition for fluid pressures and velocities at the jump interface, as well as the pressure fields and velocity fields of oil and water phase in the capillary.

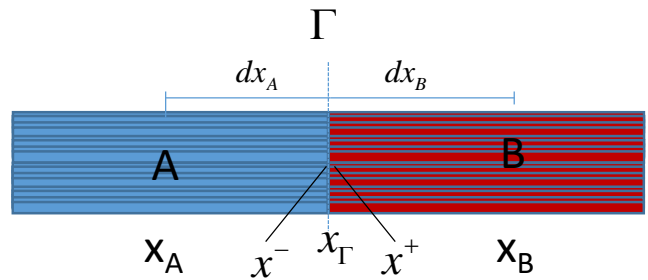

Fig.5 Diagram of Piston-like Water-oil Displacement

Table 2 Capillary Pressure Curve and Relative Permeability Curve of the Capillary Bundle

| | |
|---|---|
| Capillary Pressure, MPa | $P_c = (1-s_w)^{1.5}$ |
| Water Relative Permeability | $k_{rw} = 0.8 \times s_w^2$ |
| Oil Relative Permeability | $k_{row} = (1-s_w)^3$ |

According to the equations of capillary pressure curve and relative permeability curve, we can obtain that:

$$f_w \frac{dP_c}{ds_w} = \frac{15 s_w^2 \sqrt{1-s_w}}{4 s_w^3 - 52 s_w^2 + 12 s_w - 4} \tag{36}$$

The formula above are visualized in Fig.6

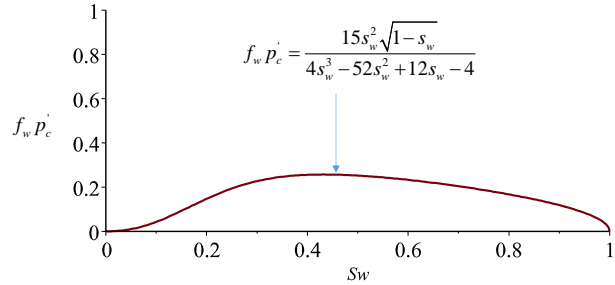

Fig.6 The plot of $f_w \dfrac{dP_c}{ds_w} - s_w$ in the Capillary Bundle

Substitute Formula (36) and other known data into the Formulas (23)-(24), via numerical integral, and get:

$$P_n^- - P_n^+ = -0.16736, \quad P_w^- - P_w^+ = 0.08264 \tag{37}$$

By virtue of Formula (15), we can determine the total pressure of Point A and B, where $P(x=x_A) = 2.16736$ MPa and $P(x=x_B) = 1.9$ MPa. Substitute them into Formula (15) to solve the Darcy velocities of water and oil: $v_w^- = v_o^+ = v_t = 1.68$ m3/day. The velocities fields in the capillary bundle are similar to that shown in Fig.4. From Formula (15), the fields of both water pressure and oil pressure can be concluded, which is as shown in Fig 7 .At the Point A side of the flood front, the water saturation is 100%, so the capillary pressure is zero and Its oil pressure is equal to its water pressure; while on the Point B side, as the water saturation is 0, capillary pressure is -0.25 MPa. The water pressure difference across the flood front is $\Delta P_w|_\Gamma = P_w^- - P_w^+ = 0.08264$ MPa, and the oil pressure difference across the interface is $\Delta P_n|_\Gamma = P_n^- - P_n^+ = -0.16736$ MPa. In a word , at the flood front, the oil pressures and water pressure is not continuous.

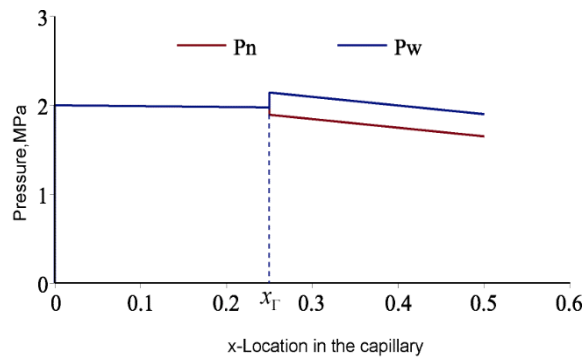

Fig.7 Fluid pressures in the Capillary Bundle

**4.3 Case 3. Non-piston Water to Oil Displacement in the Rock Core of One-dimensional Level**

As shown in Fig 8, there is a horizontal rock core, water displaces oil in the non-piston way; flood front is presented as $\Gamma$, whose coordinates $x_\Gamma = 0.5$m; capillary pressure curve is as shown in Table 3; relative permeability function of water saturation is known (same as Table 2); the absolute permeability $K = 0.1$D; water viscosity is 0.5 cp; oil viscosity is 5 cp; saturation distribution is as shown in Fig 8, Table 3, where the flood front saturation $s_{wf} = 0.3104$; seepage velocity at water injecting hole is 5 m/day and water phase pressure is 2 MPa. With losing sight of the gravity, it is required to compute the pressure fields and velocity fields of oil and water within the rock core.

Table 3 Capillary Pressure Function of Water Saturation in the Rock Core

| Capillary Pressure, MPa | $P_c(s_w) = 0.01\left(e^{4(1-s_w)^{1.2}} - 1\right)$ |
| --- | --- |

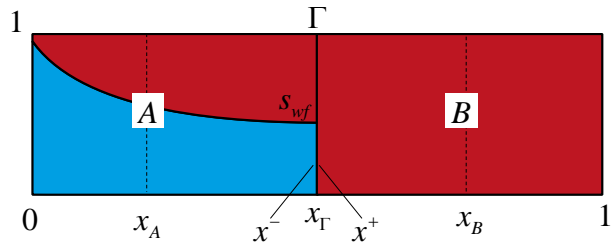

Fig.8 Oil and Water Saturation Distribution at A Certain Moment of Water Driving Oil in the Horizontal Rock

At first, calculate the value of function of $Y(s_w) = f_w(s_w)\dfrac{dP_c(s_w)}{ds_w}$ with response to $s_w$ within the range of $(0,1)$, and get the curve of $\int_0^{s_w} f_w \dfrac{dP_c}{ds_w} ds \sim s_w$, as shown in Fig 9.

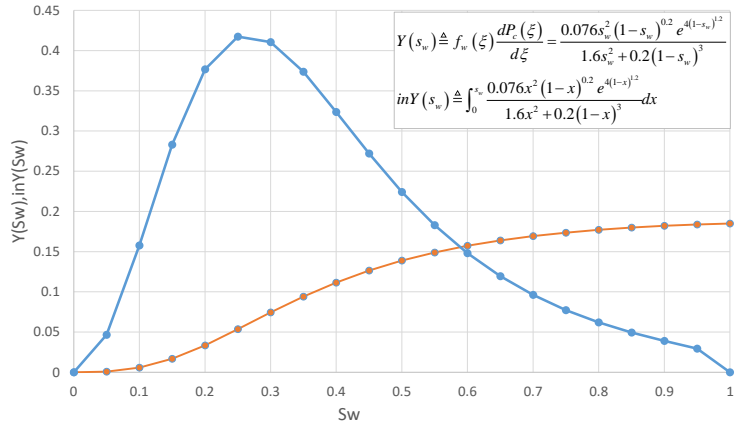

Fig.9 $f_w \dfrac{dP_c}{ds_w}, \int_0^{s_w} f_w \dfrac{dP_c}{ds} d \sim s_w$ of the Rock Core

Next, according to the Formulas (15) and all known conditions, calculate the total pressure field in the rock core. The result is shown in Figure 10, which displays that the total pressure distributes continuously inside the rock core and gradually reduces in the flowing direction.

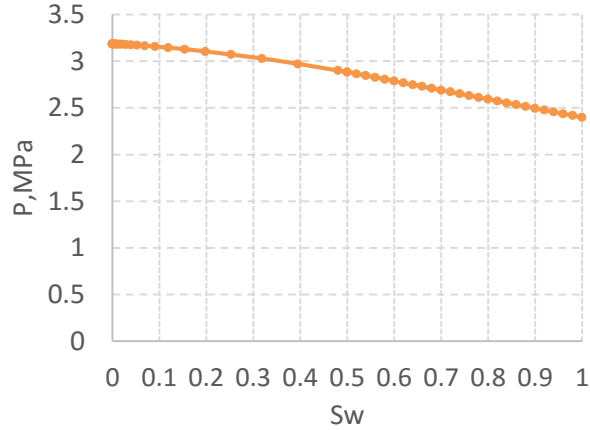

Fig.10 Total Pressure Field of the Rock Core

Then, substitute the total pressure into the Formula (14) and (9) to obtain the oil pressure and water pressure, as shown in Fig 11. At the flood front (x = 0.5 m), the oil and water pressure changes sharply. We have

$$P_w^- - P_w^+ = 0.4573 \text{ MPa}; \quad P_n^- - P_n^+ = -0.07869 \text{ MPa}. \tag{38}$$

Which prove again that oil pressure and water pressure are discontinuous at the flood front.

Finally, according to Formulas (15), (25)~(26), the Darcy velocity fields of water and oil can be obtained, as shown in Fig 12.

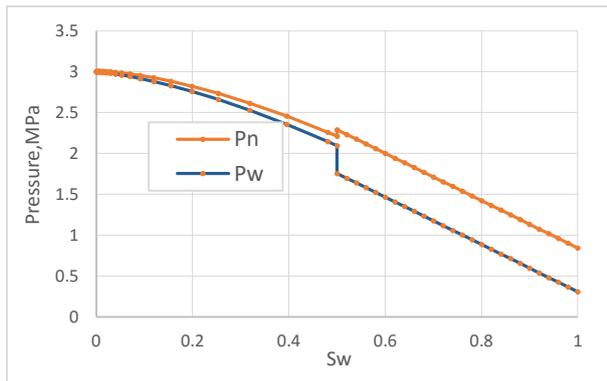
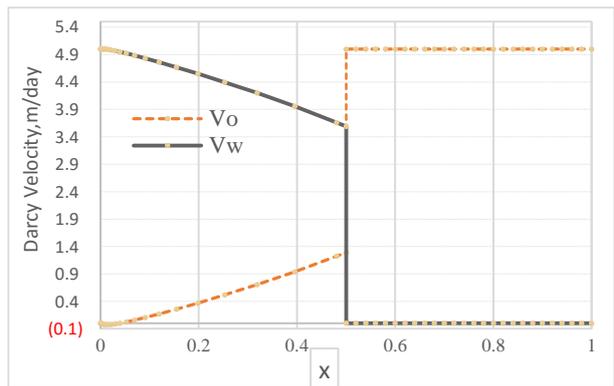

Fig.11 Fluid Pressure in the Rock Core      Fig.12 Darcy Velocity in the Rock Core

（the water injection rate is 5m/day）   （the water injection rate is 5m/day）

Taking a further step reducing the water injection rate in Case 3 from 5 m/day to 1.68 m/day, and the results about the pressures and seepage velocity can be found in Figs 13-14. The influence of the injection rate will be discussed in the next section.

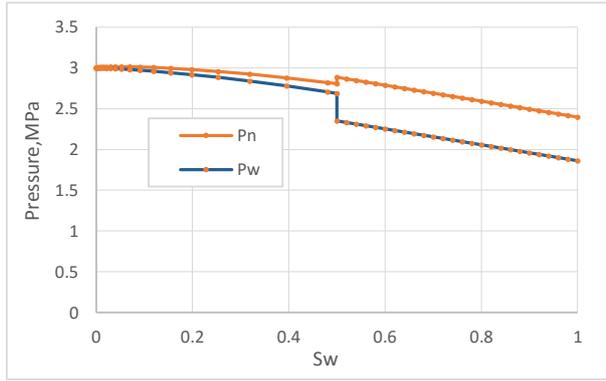
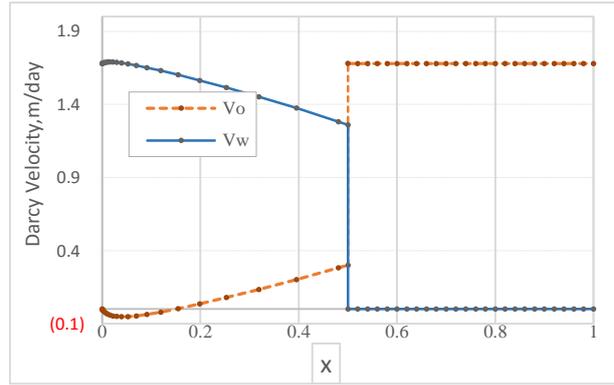

Fig.13 Fluid Pressure in the Rock Core     Fig.14 Darcy Velocity in the Rock Core

（the water injection rate is 1.6m/day）     （the water injection rate is 1.6m/day）

**4.4 Results and Discussion**

(1) It is shown in all Cases, oil pressure or water pressures jump at the flood front, so do the fluid velocities, which have verified the interface condition model in this paper. Especially, Case 1 is a very good example to verify the discontinuity of fluid pressure.

(2) In contrast of the pressure fields of Case 1 and that of Case 2, we can find that although their capillary pressure differences at $\Gamma$ are the same (0.25MPa), the fluid pressures jump at the flood front $\Gamma$ are different, which justifies that the fluid pressure jump at the flood front is related not only to the capillary pressure value, but to the whole capillary pressure curve.

(3) Because $f_w, \dfrac{dp_c}{ds_w}$ is generally simplified as the function of fluids saturation, in view of Formulas (23) and (24), while fluid distributions are determined in the certain porous media, the difference of water pressure and oil pressure on both side of the flood front is identical. It can be testified by comparing Fig 10 and Fig 12.

(4) Comparing Fig 11 with Fig 13, we can find the difference of Darcy velocities on both side of the flood front relates not only to the fluid distribution, capillary pressure curve, relative permeability curve, but to the displacing speed.

(5) In Case 3, the calculated results listed in Table 4 show that near the end of water injection, oil velocity is negative. That is to say, in these areas oil flows in the opposite direction of that of water flooding. In contrast of Fig 11 and Fig 13, it can be concluded that the scope of the reverse flow region increases at the reduction of velocity.

## 5. Conclusions

(1) The paper, via theoretical analysis and examples, has proven that the pressure and Darcy velocity of each fluid at flood front are discontinuous, which is opposite to the conventional interface condition model. For sake of easy distinction,

All its interface conditions are collectively named as Jump Pressures and Velocities Conditions Model (JPVCM).

(2) With a reference to the proof process of the total pressure satisfying continuity conditions in this paper, we hold the opinion that: in accordance with the multiphase seepage theory, if the Darcy velocity of each fluid is considered continuous, then the pressure of the fluid must be continuous at the flood front.

(3) The new interface condition model can also be expressed as the total pressure and total Darcy velocity which are continuous at the flood front under the condition that the fluids and porous medium are uncompressible. But this conclusion is untenable if their compressibility is considered.

(4) It is worth noting that this paper studies the interface condition at discontinuous interface in the macro scale. The pressure that appears in Darcy's law is an average of the physical pressure of the fluid in the pores over a representative elementary volume and the continuum pressure is discontinuous across the interface (Beatrice Riviere, 2013). But at the macro-scale, the continuum pressure is discontinuous across the jump interface.

(5) When the fluid is continuously distributed and so is the porous medium, the JPVCM still holds to be reasonable; namely, it is also suitable for interface condition of continuous interface. Even as for the discontinuous interface caused by the mutational change of porous medium, it is still suitable.

(6) JPVCM can be used directly to establish the numerical flux through the interface between two grid cells. It can be used to establish the reservoir simulation model directly.

## Acknowledgments

This work is supported by the National Science and Technology Major Project of the Ministry of Science and Technology of China 2011ZX05060-004,2011ZX0514-004,2011ZX05049-004 and National Natural Science Foundation of China (Grant No. 51474179)

Table 4 Case 3. Seepage Parameters at Different Locations in the Rock Core

| Position m | Sw | Pc MPa | $dP/dx$ MPa/m | $\int_0^s f_w(\xi)\frac{dP_c(\xi)}{d\xi}d\xi$ MPa | $P_n - \int_0^s f_w(\xi)\frac{dP_c(\xi)}{d\xi}d\xi$ MPa | $P_n$ MPa | $P_w$ MPa | $v_n$ m/day | $v_w$ m/day |
|---|---|---|---|---|---|---|---|---|---|
| 0 | 1 | 0 | 0 | -0.18496 | 3.18496 | 3 | 3 | 0 | 5 |
| 0.00182 | 0.87584 | 0.00387 | -0.00079 | -0.18109 | 3.18417 | 3.00308 | 2.99921 | -0.00881 | 5.00259 |
| 0.00314 | 0.8448 | 0.00534 | -0.00144 | -0.17963 | 3.18353 | 3.0039 | 2.99856 | -0.01297 | 5.00322 |
| 0.005 | 0.81376 | 0.00703 | -0.00241 | -0.17794 | 3.18255 | 3.00461 | 2.99758 | -0.01714 | 5.0032 |
| 0.00754 | 0.78272 | 0.00897 | -0.00386 | -0.176 | 3.1811 | 3.0051 | 2.99613 | -0.02075 | 5.00203 |
| 0.01097 | 0.75168 | 0.01121 | -0.00597 | -0.17377 | 3.17899 | 3.00523 | 2.99402 | -0.02308 | 4.9991 |
| 0.01553 | 0.72064 | 0.01377 | -0.009 | -0.17121 | 3.17596 | 3.00475 | 2.99098 | -0.02324 | 4.99363 |
| 0.02151 | 0.6896 | 0.01671 | -0.01334 | -0.16829 | 3.17162 | 3.00333 | 2.98662 | -0.02013 | 4.98465 |
| 0.02933 | 0.65856 | 0.02009 | -0.01952 | -0.16495 | 3.16544 | 3.0005 | 2.98041 | -0.01241 | 4.97094 |
| 0.03949 | 0.62752 | 0.02397 | -0.02831 | -0.16112 | 3.15666 | 2.99554 | 2.97157 | 0.00157 | 4.95095 |
| 0.05263 | 0.59648 | 0.02843 | -0.04079 | -0.15675 | 3.14418 | 2.98743 | 2.959 | 0.02385 | 4.92277 |
| 0.06957 | 0.56544 | 0.03355 | -0.0585 | -0.15176 | 3.12646 | 2.9747 | 2.94115 | 0.05693 | 4.88397 |
| 0.09131 | 0.5344 | 0.03945 | -0.08362 | -0.14608 | 3.10134 | 2.95526 | 2.91581 | 0.10388 | 4.83152 |
| 0.11906 | 0.50336 | 0.04624 | -0.11916 | -0.13962 | 3.0658 | 2.92617 | 2.87993 | 0.16848 | 4.76164 |
| 0.15424 | 0.47232 | 0.05407 | -0.16923 | -0.13231 | 3.01573 | 2.88342 | 2.82935 | 0.25527 | 4.66969 |
| 0.19836 | 0.44128 | 0.0631 | -0.23919 | -0.12408 | 2.94577 | 2.82169 | 2.75859 | 0.36965 | 4.55008 |
| 0.25287 | 0.41024 | 0.07353 | -0.33562 | -0.11486 | 2.84935 | 2.73448 | 2.66095 | 0.51775 | 4.39626 |
| 0.31869 | 0.3792 | 0.08558 | -0.46562 | -0.10464 | 2.71934 | 2.6147 | 2.52912 | 0.70606 | 4.20105 |
| 0.39557 | 0.34816 | 0.09952 | -0.63494 | -0.09345 | 2.55002 | 2.45657 | 2.35705 | 0.94038 | 3.95736 |
| 0.48092 | 0.31712 | 0.11565 | -0.844 | -0.08139 | 2.34097 | 2.25957 | 2.14392 | 1.22342 | 3.6599 |
| 0.5* | 0.3104 | 0.11946 | -0.8937 | -0.07869 | 2.29127 | 2.21258 | 2.09311 | 1.2908 | 3.58827 |
| 0.50001* | 0 | 0.53598 | -0.00003 | 0 | 2.29124 | 2.29124 | 1.75525 | 5 | 0 |
| 0.60001 | 0 | 0.53598 | -0.28938 | 0 | 2.00189 | 2.00189 | 1.4659 | 5 | 0 |
| 0.70001 | 0 | 0.53598 | -0.57873 | 0 | 1.71254 | 1.71254 | 1.17655 | 5 | 0 |
| 0.80001 | 0 | 0.53598 | -0.86808 | 0 | 1.42319 | 1.42319 | 0.8872 | 5 | 0 |
| 0.90001 | 0 | 0.53598 | -1.15743 | 0 | 1.13384 | 1.13384 | 0.59785 | 5 | 0 |
| | 0 | 0.11946 | -0.8937 | 0 | 0.84452 | 0.84452 | 0.30853 | 5 | 0 |